\documentclass[fleqn,twoside]{article}
\usepackage[headings]{espcrc2}

\readRCS
$Id: espcrc2.tex,v 1.2 2004/02/24 11:22:11 spepping Exp $
\usepackage{graphicx}

\def\bea{\begin{eqnarray}}
\def\eea{\end{eqnarray}}
\def\be{\begin{equation}}
\def\ee{\end{equation}}

\def\d{\partial}
\def\eqref#1{(\ref{#1})}
\def\a{\alpha}

\def\bra{\langle}
\def\ket{\rangle}
\def\e{{\rm e}}
\def\tr{{\rm tr}}

\newcommand{\bi}{\begin{itemize}}
\newcommand{\ei}{\end{itemize}}

\title{Numerical evidence for the Maldacena conjecture
in two-dimensional $\mathcal{N}=(8,8)$ super Yang--Mills theory%
\thanks{Preprint UMN-D-05-3,                         
to appear in the proceedings of the Workshop on 
Light-Cone QCD and Nonperturbative Hadron Physics,
Cairns, Australia, July 7-15, 2005.}%
}

\author{
J.R. Hiller%
\address{Department of Physics,
University of Minnesota-Duluth,
Duluth, MN 55812 USA}%
}
       
\runtitle{Numerical evidence for the Maldacena conjecture}
\runauthor{J.R. Hiller}

\begin{document}

\begin{abstract}
The $\mathcal{N}=(8,8)$ super Yang--Mills theory in $1+1$ dimensions is solved
at strong coupling to directly confirm the predictions of supergravity
at weak coupling.  The calculations are done in the large-$N_c$
approximation using Supersymmetric Discrete Light-Cone Quantization.
The stress-energy correlator is obtained as a function of the separation
$r$; for intermediate values of $r$, the correlator behaves in a manner
consistent with the $1/r^5$ behavior predicted by weak-coupling
supergravity.
\vspace{1pc}
\end{abstract}

\maketitle

\section{INTRODUCTION}

The conjectured correspondences~\cite{Maldacena1,Maldacena2} between certain 
string theories and supersymmetric Yang--Mills (SYM) theories at large-$N_c$
can be tested directly if one is able to solve an SYM theory at
strong coupling.  The solution can then be compared to a small-curvature
supergravity approximation to the corresponding string theory.  This
approach requires a nonperturbative technique for SYM theories, and 
one has been developed over the past several years as a supersymmetric
form~\cite{Sakai,SDLCQreview} of discrete light-cone 
quantization~\cite{PauliBrodsky,DLCQreview}, known as 
SDLCQ.

The SDLCQ approach is a Hamiltonian formulation in a Fock basis using
light-cone coordinates~\cite{Dirac}.  The choice of coordinates allows
for a simple vacuum and a consistent Fock expansion.  The momentum-space
wave functions in the different Fock sectors then satisfy coupled integral 
equations which can be discretized to obtain a matrix representation of the 
bound-state eigenvalue problem.  The discretization is actually done at 
the level of second-quantized operators expanded in momentum modes;
however, this ultimately yields the same matrix eigenvalue problem.

The quantity which we compute in order to test a correspondence is
a two-point correlator of the stress-energy tensor.  
The correlator is computed
in SDLCQ by inserting a sum over the eigenstates obtained in
the matrix diagonalization. To keep the
calculation manageable, we consider the case of $\mathcal{N}=(8,8)$
SYM theory in two dimensions which corresponds to a particular
type IIB string theory~\cite{Maldacena2}.  The supergravity 
approximation to the correlator is discussed in \cite{JHEP}.
The potential for an SDLCQ check of this correspondence was 
explored there and in subsequent work~\cite{subsequent}, but the 
numerical resolution available in this earlier work
was insufficient for a true test. We have now reached a
resolution sufficient to provide evidence of consistency
in the correspondence~\cite{newwork}.

An outline of the remainder of the paper is as follows.
Section~\ref{sec:SDLCQ} describes the SDLCQ method.  The
particular SYM theory is described in Sec.~\ref{sec:SYM},
and the calculation of the correlator is formulated in
Sec.~\ref{sec:Correlator}.  For comparison purposes,
we consider both $\mathcal{N}=(8,8)$ and $\mathcal{N}=(2,2)$ 
theories.  The results are presented
and discussed in Sec.~\ref{sec:Results}.  
Some additional discussion is provided in 
Sec.~\ref{sec:Conclusions}.

\section{\mbox{SUPERSYMMETRIC DISCRETE} LIGHT-CONE QUANTIZATION}
\label{sec:SDLCQ}

We use light-cone coordinates as suggested by Dirac~\cite{Dirac}.
The time coordinate is $x^+=(t+z)/\sqrt{2}$, and the space
coordinate is $x^-\equiv (t-z)/\sqrt{2}$.  The conjugate
variables are the light-cone energy $p^-=(E-p_z)/\sqrt{2}$
and momentum $p^+\equiv (E+p_z)/\sqrt{2}$.  The mass-shell 
condition $p^2=m^2$ then yields $p^-=\frac{m^2}{2p^+}$

The discretization proposed by Pauli and Brodsky~\cite{PauliBrodsky},
known as discrete light-cone quantization (DLCQ)~\cite{DLCQreview},
is based on the imposition of periodic boundary conditions
on a light-cone box $-L<x^-<L$.  This yields a discrete grid
in momentum space where individual momenta are specified by
$p_i^+=\frac{\pi}{L}n_i$, with $n_i$ a positive integer.
For fixed total momentum $P^+$, the limit $L\rightarrow\infty$ 
is exchanged for a limit in terms of the integer resolution
$ K\equiv\frac{L}{\pi}P^+$.  Since the individual momenta are
strictly positive, $K$ is the upper limit on the number of
particles allowed by the discretization.
Integrals are replaced by discrete sums
\be 
\int dp^+ f(p^+)\simeq  \frac{\pi}{L} \sum_n f(n\pi/L), 
\ee
and Dirac delta functions become Kronecker deltas
\be 
\delta(p^+-p^{'+})\rightarrow\frac{L}{\pi}\delta_{nn'}. 
\ee

Supersymmetric DLCQ (SDLCQ)~\cite{SDLCQreview} is 
constructed~\cite{Sakai} to maintain the supersymmetry algebra
\bea
\{Q^+,Q^+\}&=&2\sqrt{2}P^+, \\
\{Q^-,Q^-\}&=&2\sqrt{2}P^-, \nonumber \\
\{Q^+,Q^-\}&=&-4P_\perp.  \nonumber
\eea
This algebra is satisfied explicitly by first
discretizing the supercharge $Q^-$ and computing
\be 
P_{\rm SDLCQ}^-=\frac{1}{2\sqrt{2}}\left\{Q^-,Q^-\right\}.
\ee
In ordinary DLCQ, the $P^-$ operator is discretized directly
and the supersymmetry algebra is not satisfied, except in
the limit of infinite resolution.  By preserving the
supersymmetry algebra, SDLCQ preserves supersymmetry in
the spectrum, even at finite resolution.

\section{SUPER YANG--MILLS THEORIES}
\label{sec:SYM}

The ${\cal N}$=(8,8) SYM theory is obtained by reducing
${\cal N}=1$ SYM theory from ten to two dimensions.
The action in light-cone gauge ($A_-=0$) is~\cite{Antonuccio:1998tm}
\bea
\lefteqn{S^{LC}_{1+1}=\int dx^+ dx^- \tr \Bigg[ \d_+ X_I \d_-X_I}&& \\
 && +i\theta^T_R \d^+\theta_R+i\theta^T_L\d^-\theta_L \nonumber \\
 &&+\frac 12 (\d_-A_+)^2+gA_+J^+  \nonumber \\
 &&+\sqrt 2 g\theta^T_L\beta_I
 [X_I,\theta_R]+\frac {g^2}4 [X_I,X_J]^2 \Bigg]. \nonumber
\eea
Here the $X_I$, with $I=1,\ldots,8$, are
the scalar remnants of the transverse components of the
ten-dimensional gauge field $A_{\mu}$.
The two-component spinor fields
$\theta_R$ and $\theta_L$ are remnants of the right-moving and left-moving
projections of the original sixteen-component spinor.
We also have the current $J^+=i[X_I,\d_-X_I]+2\theta^T_R\theta_R$
and two matrices $\beta_1\equiv\sigma_1$, $\beta_2\equiv\sigma_3$.
The supercharges for this theory can be obtained by dimensionally reducing
the ten-dimensional supercurrent, which yields
\bea
Q^-_{\a}&=&g\int dx^- \tr\left( -2^{3/4}J^+\frac1{\d_-}u_{\a}\right. \\
&&\left.+2^{-1/4}i[X_I,X_J](\beta_I\beta_J)_{\a\eta
}u_{\eta}\right), \nonumber
\eea
where $\a,\eta=1,\ldots,8$
and the $u_{\a}$ are the components of $\theta_R$.
The mode expansions of the dynamical fields are
\bea
\lefteqn{X_{Ipq}(x^-)=\frac 1{\sqrt{2\pi}}\int_0^{\infty}
\frac {dk^+}{\sqrt{2k^+}}} && \\
&&\times[A_{Ipq}(k^+)e^{-ik^+x^-}
 +A^{\dag}_{Iqp}(k^+)e^{ik^+x^-}],\nonumber
\eea
and
\bea
\lefteqn{u_{\a pq}(x^-)=\frac 1{\sqrt{2\pi}}\int_0^{\infty}
   \frac {dk^+}{\sqrt{2}}} && \\
&&\times[B_{\a pq}(k^+)e^{-ik^+x^-}
   +B^{\dag}_{\a qp}(k^+)e^{ik^+x^-}],\nonumber
\eea
where $p,q=1,2,\ldots,N_c$.

The operators $A$ and $B$ satisfy the (anti)commutation relations
\bea
\lefteqn{[A_{Ipq}(k^+),A^{\dag}_{Jrs}(k^{'+})]}&& \\
  && =\delta_{IJ}\delta_{pr}\delta_{qs}\delta(k^+-k^{'+}),  \nonumber \\
\lefteqn{\{B_{\a pq}(k^+),B^{\dag}_{\beta rs}(k^{'+})\}} && \nonumber \\
   && =\delta_{\a\beta}\delta_{pr}\delta_{qs}\delta(k^+-k^{'+}). \nonumber
\eea
In the discrete approximation, we
rescale the annihilation operators as
\bea
\sqrt{\frac{L}{\pi}}a(k) = A(k^+ = \frac{\pi k}{L}), \\
\sqrt{\frac{L}{\pi}}b(k) =  B(k^+ = \frac{\pi k}{L}). \nonumber
\eea
We then have
\bea
[a_{Ipq}(k),a^{\dag}_{Jrs}(k')]
   &=&\delta_{IJ}\delta_{pr}\delta_{qs}\delta_{kk'},  \\
\{b_{\a pq}(k),b^{\dag}_{\beta rs}(k')\}
   &=&\delta_{\a\beta}\delta_{pr}\delta_{qs}\delta_{kk'},
\eea
and
\bea
X_{Ipq}(x^-)&=&\frac 1{\sqrt{2\pi}}\sum_{k=1}^{\infty}
\frac {1}{\sqrt{2k}}[a_{Ipq}(k)e^{-i\frac{\pi}{L}kx^-}\nonumber \\
 && +a^{\dag}_{Iqp}(k^+)e^{i\frac{\pi}{L}kx^-}] \\
u_{\a pq}(x^-)&=&\frac 1{\sqrt{2L}}\sum_{k=1}^{\infty}
   \frac {1}{\sqrt{2}}[b_{\a pq}(k)e^{-i\frac{\pi}{L}kx} \nonumber\\
  && +b^{\dag}_{\a qp}(k)e^{i\frac{\pi}{L}kx^-}]. \nonumber
\eea

The presence of extended supersymmetry means that there are eight
different Hamiltonians $P_\a^-=\{Q_\a^-,Q_\a^-\}/2\sqrt{2}$, any one of 
which can be diagonalized.  There exist unitary transformations between
them to guarantee that the spectrum is the same for all.  We will
work with $P_8^-$.

This Hamiltonian can be block diagonalized by taking advantage of
various symmetries~\cite{newwork}.  The generators of these
symmetries are given in Table~\ref{tab:Symmetries}.
\begin{table*}
\caption{Generators of symmetries for the Hamiltonian $P_8^-$.}
\label{tab:Symmetries}
\begin{tabular}{rrrrrrrrrrrrrrrr}
\hline
   & $\!\!a_1\!\!$& $\!\!a_2\!\!$& $\!\!a_3\!\!$& $\!\!a_4\!\!$& $\!\!a_5\!\!$& $\!\!a_6\!\!$& $\!\!a_7\!\!$& $\!\!a_8\!\!$
   & $\!\!b_1\!\!$& $\!\!b_2\!\!$& $\!\!b_3\!\!$& $\!\!b_4\!\!$& $\!\!b_5\!\!$& $\!\!b_6\!\!$& $\!\!b_7\!\!$\\
\hline
$\!\!1\!\!$  & $\!\!a_1\!\!$& $\!\!a_8\!\!$& $\!\!-a_5\!\!$& $\!\!-a_4\!\!$& $\!\!-a_3\!\!$& $\!\!a_6\!\!$& $\!\!-a_7\!\!$& $\!\!a_2\!\!$
   & $\!\!b_1\!\!$& $\!\!b_4\!\!$& $\!\!-b_3\!\!$& $\!\!b_2\!\!$& $\!\!b_7\!\!$& $\!\!-b_6\!\!$& $\!\!b_5\!\!$\\
$\!\!2\!\!$  & $\!\!a_2\!\!$& $\!\!a_1\!\!$& $\!\!-a_5\!\!$& $\!\!-a_6\!\!$& $\!\!-a_3\!\!$& $\!\!-a_4\!\!$& $\!\!-a_8\!\!$& $\!\!-a_7\!\!$
   & $\!\!b_4\!\!$& $\!\!b_3\!\!$& $\!\!b_2\!\!$& $\!\!b_1\!\!$& $\!\!b_5\!\!$& $\!\!-b_6\!\!$& $\!\!-b_7\!\!$\\
$\!\!3\!\!$  & $\!\!a_2\!\!$& $a_1\!\!$& $\!\!-a_6\!\!$& $\!\!a_8\!\!$& $\!\!a_7\!\!$& $\!\!-a_3\!\!$& $\!\!a_5\!\!$& $\!\!a_4\!\!$
   & $\!\!b_1\!\!$& $\!\!-b_2\!\!$& $\!\!b_6\!\!$& $\!\!b_5\!\!$& $\!\!b_4\!\!$& $\!\!b_3\!\!$& $\!\!-b_7\!\!$\\
$\!\!4\!\!$  & $\!\!-a_1\!\!$& $\!\!-a_2\!\!$& $\!\!-a_3\!\!$& $\!\!-a_4\!\!$& $\!\!-a_5\!\!$& $\!\!-a_6\!\!$& $\!\!-a_7\!\!$& $\!\!-a_8\!\!$
   & $\!\!b_1\!\!$& $\!\!b_2\!\!$& $\!\!b_3\!\!$& $\!\!b_4\!\!$& $\!\!b_5\!\!$& $\!\!b_6\!\!$& $\!\!b_7\!\!$\\
$\!\!5\!\!$  & $\!\!a_1\!\!$& $\!\!a_2\!\!$& $\!\!a_3\!\!$& $\!\!-a_4\!\!$& $\!\!-a_5\!\!$& $\!\!a_6\!\!$& $\!\!-a_7\!\!$& $\!\!-a_8\!\!$
   & $\!\!b_1\!\!$& $\!\!b_2\!\!$& $\!\!-b_3\!\!$& $\!\!-b_4\!\!$& $\!\!-b_5\!\!$& $\!\!-b_6\!\!$& $\!\!b_7\!\!$\\
$\!\!6\!\!$  & $\!\!-a_1\!\!$& $\!\!a_2\!\!$& $\!\!a_3\!\!$& $\!\!-a_4\!\!$& $\!\!a_5\!\!$& $\!\!-a_6\!\!$& $\!\!-a_7\!\!$& $\!\!a_8\!\!$
   & $\!\!b_1\!\!$& $\!\!-b_2\!\!$& $\!\!b_3\!\!$& $\!\!-b_4\!\!$& $\!\!-b_5\!\!$& $\!\!b_6\!\!$& $\!\!-b_7\!\!$\\
$\!\!7\!\!$  & $\!\!a_1\!\!$& $\!\!-a_2\!\!$& $\!\!a_3\!\!$& $\!\!a_4\!\!$& $\!\!-a_5\!\!$& $\!\!-a_6\!\!$& $\!\!-a_7\!\!$& $\!\!a_8\!\!$
   & $\!\!-b_1\!\!$& $\!\!b_2\!\!$& $\!\!b_3\!\!$& $\!\!-b_4\!\!$& $\!\!b_5\!\!$& $\!\!-b_6\!\!$& $\!\!-b_7\!\!$\\
\hline
\end{tabular}
\end{table*}
The block diagonalization significantly reduces the computational load
for diagonalization.

The ${\cal N}$=(2,2) SYM theory is obtained through reduction of
the ${\cal N}$=1 SYM theory from four to two dimensions~\cite{N22}.
The action is the same as for ${\cal N}$=(8,8), except that
the indices run from 1 to 2 instead of 1 to 8.
Just as there are fewer dynamical fields, there are also fewer symmetries.
The smaller number of fields allows calculations at
higher resolution; we have reached $K=14$ for the (2,2) theory
vs $K=11$ for the (8,8) theory.
However, there is no conjecture of correspondence or
any separate estimate of the correlator's behavior
for the $(2,2)$ theory.


\section{STRESS-ENERGY CORRELATOR}
\label{sec:Correlator}

The stress-energy correlation function that we compute is
\be
F(x^-,x^+)  \equiv\bra T^{++}(x)T^{++}(0)\ket.
\ee
For the string theory corresponding 
to two-dimensional ${\cal N}$=(8,8) SYM theory, $F$
can be calculated on the string-theory side
in a weak-coupling super-gravity approximation.
Its behavior for intermediate separations $r\equiv\sqrt{2x^+x^-}$ 
is~\cite{JHEP}
\be
F(x^-,x^+)=\frac {N_c^{\frac 32}}{g_{\rm YM}r^5}.
\ee
We will compute $F$ in SYM theory
and compare, considering both ${\cal N}=(8,8)$ and $(2,2)$
theories.

We fix the total momentum $P^+=P_-$ and compute
the Fourier transform, which can be expressed
in a spectral decomposed form as~\cite{JHEP}
\bea
\lefteqn{\tilde F(P_-,x^+)} && \\
    &&   =\frac 1{2L}\bra T^{++}(P_-,x^+) T^{++}(-P_-,0)\ket \nonumber \\
   &&=\sum_i \frac 1{2L}\bra 0|T^{++}(P_-,0)|i\ket\e^{-iP^i_+x^+} \nonumber \\
       && \rule{0.75in}{0mm}  \times  \bra i|T^{++}(-P_-,0)|0\ket. \nonumber
\eea
The position-space form is recovered by the inverse transform,
with respect to $P_-=K\pi/L$. 
The continuation to Euclidean space is made by 
taking $r$ to be real.  This yields 
\bea
\lefteqn{F(x^-,x^+)=\sum_i \Big|\frac L{\pi}\bra 0|T^{++}(K)|i\ket\Big|^2} && \\
  && \times \left(\frac {x^+}{x^-}\right)^2
   \frac{M_i^4 K_4(M_i \sqrt{2x^+x^-})}{8\pi^2K^3}. \nonumber
\eea

The stress-energy operator $T^{++}$ is given by 
\bea
\lefteqn{T^{++}(x^-,x^+) =\tr \left[(\d_- X^I)^2\right.} && \\
&& \left.+\frac 12(iu^{\a}\d_-u^{\a}-i(\d_-u^{\a})u^{\a})\right]. \nonumber
\eea
In terms of the discretized creation operators, we find 
\bea
\lefteqn{T^{++}(-K)|0\ket} && \\
&& =\frac {\pi}{2L}\sum_{k=1}^{K-1} 
    \left[-\sqrt{k(K-k)} a^{\dag}_{Iij}(K-k)a^{\dag }_{Iji}(k)\right. \nonumber\\
 && +\left.\left(\frac K2-k\right)
   b^{\dag }_{\a ij}   (K-k)b^{\dag }_{\a ji}(k)\right]|0\ket. \nonumber
\eea
Thus $(L/\pi)\bra 0|T^{++}(K)|i\ket$ is independent of $L$. 
Also, only one symmetry sector contributes.

The correlator behaves like $1/r^4$ at small $r$, as can be seen 
by taking the limit to obtain
$\left(\frac {x^-}{x^+}\right)^2F(x^-,x^+)
   \sim\frac{N_c^2(2n_b+n_f)}{4\pi^2r^4}(1-1/K)$.
To simplify the appearance of this behavior, we rescale $F$ by defining
\bea
\lefteqn{f\equiv \bra T^{++}(x)T^{++}(0)\ket 
      \left(\frac{x^-}{x^+}\right)^2} &&  \\
      &&  \times \frac{4\pi^2r^4}{N_c^2(2n_b+n_f)}. \nonumber
\eea
Then $f$ is just $(1-1/K)$ for small $r$.

We compute $f$ numerically by obtaining the entire spectrum for 
small matrices and by using Lanczos iterations for large matrices.
The Lanczos technique~\cite{subsequent} generates an approximate 
tridiagonal representation of the Hamiltonian which captures the
important contributions after only a few iterations and which
is easily diagonalized to compute the sum over eigenstates.

\section{RESULTS}
\label{sec:Results}

\begin{figure}[htbp]
\begin{tabular}{c}
\includegraphics[width=5.5cm]{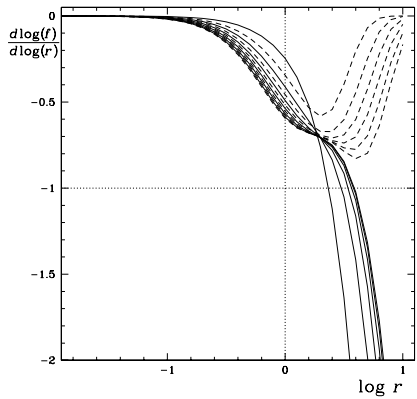} \\
(a) \\
\includegraphics[width=5.5cm]{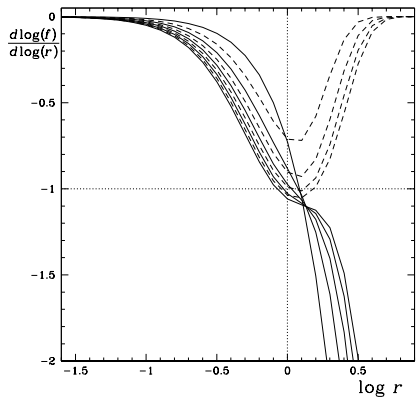} \\
(b)
\end{tabular}
\caption{Plots of the log-log derivative of the
rescaled correlator $f$ for the (a) $\mathcal{N}=(2,2)$ and (b) $(8,8)$
SYM theories.  Each curve corresponds to a different resolution $K$,
with $K$ ranging from 3 to 14 in (a) and from 3 to 11 in (b).  For odd $K$
the curves are solid, and for even $K$ they are dashed.
The separation $r$ is measured in units of $\sqrt{\pi/g^2N_c}$.}
\label{fig:loglog}
\end{figure}

The log-log derivative of the rescaled correlator $f$ is
plotted in Fig.~\ref{fig:loglog} for both SYM theories
and for a range of resolution values.
At small $r$, the graphs for different $K$ match the expected $(1-1/K)$ behavior. 
At large $r$, the behavior is different between odd and even $K$, although in
the intermediate region, the difference gets smaller as $K$ gets bigger. 
The difference in behavior at large $r$ is due to the absence of an exactly
massless state among states that contribute for $K$ odd.  For each even $K$
there is a contributing massless state, which allows the correlator to return
to the proper $1/r^4$ behavior at large $r$.  For odd $K$, the lowest 
contributing state becomes massless only in the large-$K$ limit.
In the $\mathcal{N}=(8,8)$ theory, the corresponding supergravity
solution for intermediate $r$ implies that the log-log derivative
of the rescaled correlator should be equal to $-1$.

\begin{figure}[htbp]
\begin{tabular}{c}
\includegraphics[width=5.5cm]{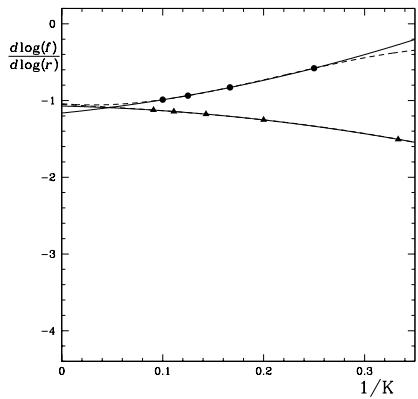} \\
(a) \\
\includegraphics[width=5.5cm]{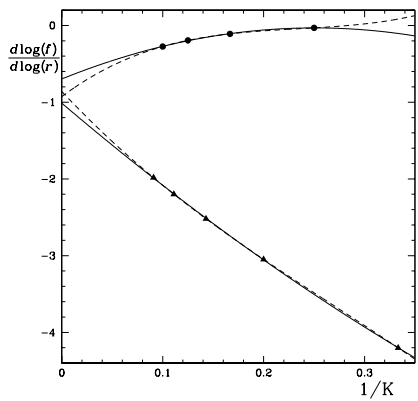} \\
(b) 
\end{tabular}
\caption{Sample extrapolations for the log-log derivative of the 
rescaled correlator $f$ in the $\mathcal{N}=(8,8)$ theory for
(a) $\log_{10}(r)=0.2$ and (b) $\log_{10}(r)=0.5$.  The lines show 
quadratic (solid) and cubic (dashed) fits to the computed points,
with the fits done separately for odd and even $K$.}
\label{fig:samples}
\end{figure}

\begin{figure}[htbp]
\begin{tabular}{c}
\includegraphics[width=5.5cm]{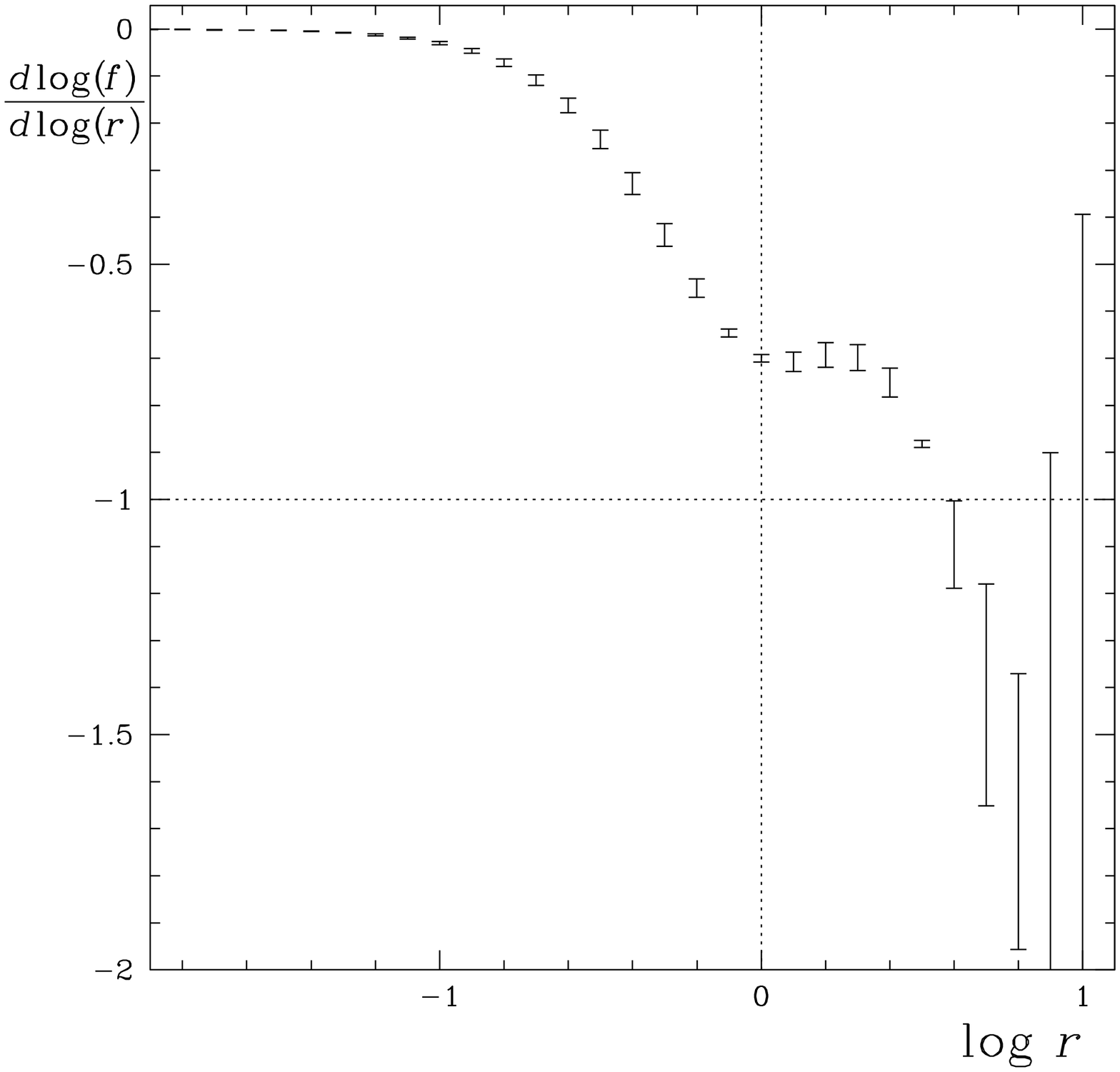} \\
(a) \\
\includegraphics[width=5.5cm]{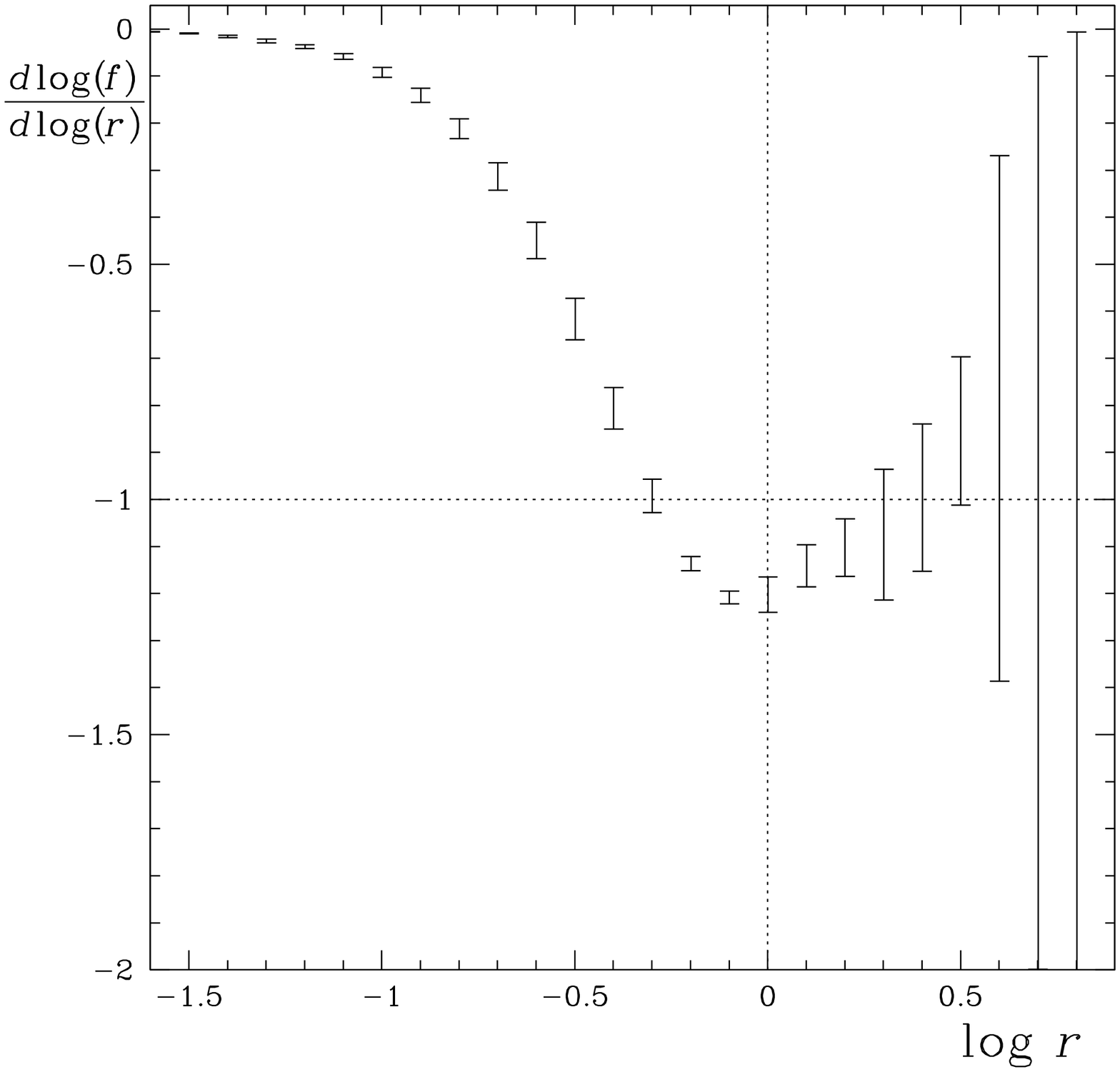} \\
(b)
\end{tabular}
\caption{Summary of extrapolations to infinite resolution
for the (a) $\mathcal{N}=(2,2)$ and (b) $(8,8)$
SYM theories.  The vertical segments represent the intervals
obtained by various choices of extrapolations, such as those
shown in Fig.~\ref{fig:samples}.}
\label{fig:extrap}
\end{figure}

For small and intermediate $r$ we can extrapolate the values of $f$ to
infinite resolution.  Typical extrapolations for the $(8,8)$ theory are 
given in Fig.~\ref{fig:samples}.
The range of values obtained with fits of different orders
to odd and even $K$ provides an estimate of the interval
within which the actual value should lie.  We display these
intervals in Fig.~\ref{fig:extrap}.
The two theories clearly differ in their behavior for intermediate $r$
and only the $(8,8)$ theory is consistent with the $1/r^5$
behavior predicted for it by the supergravity approximation
to the dual string theory.

\section{CONCLUSIONS}
\label{sec:Conclusions}

The calculations that we have done succeed in distinguishing between
theories that differ in the amount of extended supersymmetry.
Only the result for the $\mathcal{N}=(8,8)$ theory is consistent 
with a $1/r^5$ behavior and thus with the Maldacena conjecture
for this theory~\cite{Maldacena2,JHEP}.  The errors in the extrapolations
remain too large to support a claim of computing the expected
$1/r^5$ behavior, but the numerical evidence is consistent.
Additional calculations at higher resolution may soon be possible,
and these may well provide explicit confirmation of the conjecture.

\section*{ACKNOWLEDGMENTS}
The work reported here was done in collaboration with 
S. Pinsky, N. Salwen, and U. Trittmann,
and was supported in part by the US Department of Energy
and the Minnesota Supercomputing Institute.

\end{document}